\documentclass[letterpaper, 11pt]{article}
\usepackage{amsmath}  
\usepackage{amssymb}  
\usepackage{amsthm}  
\usepackage{enumitem}  
\usepackage{fullpage}  
\usepackage{graphicx}  
\usepackage{parskip}  
\usepackage{braket}  
\usepackage{qcircuit}
\usepackage{url}
\usepackage{authblk}
\usepackage{appendix}

\title{
    \vspace{-0.8in}
    \text{Sampling random quantum circuits: a pedestrian's guide}
}

\author[1]{Sean Mullane}
\affil[1]{Department of Applied Physics, Stanford University}
\date{}

\begin{document}

\maketitle

\section{Introduction}

Hello there! I'm glad you could join us. Although I'm not sure who you are, you're presumably here because you heard about the quantum supremacy experiment recently published by a collaboration from Google, NASA Ames, UC Santa Barbara, and a host of other teams from around the world \cite{NEARNAT}. You may have seen the articles declaring a ``quantum supremacy breakthrough," and had some sense that something momentous had occurred. We've finally produced a quantum computer that can outperform a classical computer! In fact, ``outperform" is a bit of a misnomer here; since the task completed by the quantum computer would take a state-of-the-art supercomputer about 10,000 years, it's more accurate to say that a classical computer could \textit{never} effectively match the quantum computer\footnote{A competing research group at IBM claimed that the sampling task completed in \cite{SUPRNAT} could be computed on a classical computer in just 2.5 days, so quantum supremacy wasn't actually achieved. I won't address their criticisms in this article, but if you're interested more details can be found here: \url{https://www.ibm.com/blogs/research/2019/10/on-quantum-supremacy/}.} \cite{SUPRNAT}. 

Given all this excitement, it's natural to wonder exactly what the ``task" was that allowed quantum supremacy to be achieved. Popular science articles covering the experiment typically mention that the quantum computer was ``performing calculations with random numbers," or something along those lines. It's difficult to find an article that goes into more detail, and it turns out there's a reason for this: the theoretical basis for the task described in \cite{SUPRNAT} is incredibly subtle, pulling concepts from random matrix theory, mathematical analysis, quantum chaos, computational complexity, and probability theory, among other fields.  

I learned this when I set out to understand the theory behind the experiment myself. What I thought would only involve a week of reading the main papers published by the Google \textit{et al} team turned into a month-long saga with me referencing, at last count, seventeen distinct textbooks and papers covering the fields described above. Not only are the basic concepts fairly complicated, the information describing those concepts is scattered to a frustrating degree. The papers \cite{NEARNAT} and \cite{SUPRNAT} written to describe the theoretical basis frequently leave out important details in key proofs, or place these details in supplements that are difficult to find.

This article is an attempt to alleviate these frustrations in others who might want to develop a good understanding of quantum supremacy. To this end, I've designed the material here to fill a gap between non-technical, popular science articles and the research papers themselves. Before beginning, you should have some technical background: I'm assuming a basic knowledge of quantum computing at the undergraduate level, and some familiarity with linear algebra, probability theory, and the very basics of complex analysis. You should already be able to explain the basics of qubits, and have an idea of how one- and two-qubit gates ($X$, $Y$, $Z$, $CNOT$, etc.) can be represented and applied to qubits in quantum circuits. It will also be helpful if you've already attempted to read sections I and II of \cite{NEARARXIV}, and have skimmed through \cite{SUPRNAT}. Don't worry if the details are still unclear after a first read, as my goal with this paper is to provide a framework to make those details clear. It will also be extremely helpful to have \cite{SUPP} nearby, as we'll be following a few of its derivations carefully in the coming sections. 

Before diving in, I want to spend a moment outlining the main questions I'll be attempting to answer, using the papers referenced above as guides. I won't be covering any details of the experimental implementation, and won't discuss the details of their results. I'm mostly concerned here with the \textit{theoretical background} of the experiment, which covers the following questions: 
    \begin{enumerate}
        \item As we'll see, the task used to demonstrate quantum supremacy was \textit{sampling from a pseudo-random quantum circuit}. What is the precise mathematical formulation of this task? 
        \item Why is it hard for a classical computer to complete this task? 
        \item How can we use this task to demonstrate a gap between the performance of a quantum computer and classical computer? In other words, how will we know when quantum supremacy is achieved, given the mathematical formulation of the sampling task? 
    \end{enumerate}
We'll use these questions as guides to explore quantum supremacy. With that, let's dive in and begin with a basic description of the sampling task. 

\section{Roadmap: where are we going?}

Before diving into the math, let's take a step back and set the stage for the experiment. We mentioned above that quantum supremacy was demonstrated by sampling from a random quantum circuit, so it'll be helpful to have a precise definition for this sampling. The overall idea is actually pretty straightforward; consider an $n$-qubit system in a $2^n$ dimensional Hilbert space $\mathcal{H}^{\otimes n}$. A quantum circuit on this qubit system can be defined by a unitary matrix $U$ that evolves an initial qubit state $|\psi_0\rangle$ to a final qubit state $|\psi\rangle = U |\psi_0\rangle$. For example, a simple circuit that applies a Hadamard gate to two qubits, each initialized to the $|0\rangle$ state, is defined by the unitary operator $U = H \otimes H$. If we measure each of the $2^n$ qubits in the computational basis after applying some operator $U$, we'll recover a bitstring $x = a_1 a_2 ... a_N$, where $a_i \in \{ 0, 1 \}$ and $N \equiv 2^n$. Note that this bitstring corresponds to a basis state $| x \rangle = |a_1 a_2 ... a_N\rangle$ for our $2^n$ dimensional system, and that the probability of our wavefunction ``collapsing" to this basis state upon measurement is given by:
    $$ p_{U}(|x\rangle) = p_{U}(x) = |\langle a_1 a_2 ... a_N | \psi \rangle |^2 = | \langle x | \psi \rangle |^2 $$
In this case, $p_{U}(|x\rangle)$ is read ``the probability, given a quantum circuit defined by the unitary operator $U$, of observing the basis state $|x\rangle$ (or bitstring $x$) upon measuring the system $|\psi\rangle$." We'll also denote this quantity by $p(x)$, when it's clear that we're referencing a specific circuit $U$. This is what we mean by ``sampling" a quantum circuit: evolve the circuit with a unitary operator $U$, and then measure each qubit in the resulting system to obtain a bitstring $x$.

Now let's consider a \textit{random} quantum circuit. If a general quantum circuit is defined by a unitary operator $U$, you might suspect that we can construct a random quantum circuit by somehow obtaining a random unitary operator, and then applying this unitary operator to our initial qubit system. This is exactly correct, although the task of obtaining a random unitary operator is no easy feat! Regardless, let's assume for now that we do have some way to obtain a random unitary operator, so we can set up a random quantum circuit; we can sample bitstrings exactly as we described above, and obtain $p_{U}(x)$ for each bitstring. If we set up $m$ copies of a quantum circuit defined by $U$, and measure the state of each circuit after evolution by $U$, we'll obtain a set of bitstrings $S = \{ x_1, x_2, ..., x_m \}$. Since each circuit is independent, the bitstrings are independent random variables, and we can write the probability of obtaining a sample $S$ as: 
    $$ \text{Pr}(S) = \prod_{x \in S} | \langle x | \psi \rangle |^2 = \prod_{x \in S} p_{U}(x) $$
Great! In theory, this is all achievable if we can implement the quantum circuit $U$ with a low enough error rate, and we'll assume for now that this is the case. Since the entire quantum supremacy concept is based on comparing the classical to quantum cases, it's natural to ask how this sampling task would adopt itself to a classical computer. We'd just need some representation of the initial state $|\psi_{0, \text{cl}}\rangle$ (a $2^n \times 1$ dimensional vector), the unitary operator $U_{\text{cl}}$ (a $2^n \times 2^n$ matrix), and the output state $|\psi_{\text{cl}}\rangle$ (the subscript ``cl" here indicates that these states and matrices were obtained with a classical simulation). We could then simulate a measurement on the output state and obtain a \textit{classically obtained} bitstring $x^{\text{cl}}$. If we repeat this simulation many times, each time sampling a random unitary matrix $U_{\text{cl}}$, we'll obtain a classical sample of bitstrings $S_{\text{cl}} = \{ x^{\text{cl}}_1, x^{\text{cl}}_2, ..., x^{\text{cl}}_m \}$, just as we did in the quantum case. As before, the probability of observing the set $S_{\text{cl}}$ is given by: 
    $$ \text{Pr}(S_{\text{cl}}) = \prod_{x^{
    \text{cl}} \in S_{\text{cl}}} | \langle x^{\text{cl}} | \psi \rangle |^2 $$
An extremely important point: \textit{we calculate the probability of $S_{\text{cl}}$ by taking the inner product of $|x^{\text{cl}}\rangle$ with the state $|\psi\rangle$ obtained from the quantum computer, \textbf{not} the state $|\psi_{\text{cl}}\rangle$ obtained from the classical simulation.}

Why is this point important? To demonstrate quantum supremacy, we're ultimately going to show that there's a \textit{measurable difference} between $\text{Pr}(S)$ and $\text{Pr}(S_{\text{cl}})$. This is based on the fact that the size of the state space for the quantum circuit is exponential in $n$, so even for a modest $n = 50$ qubits (approximately the number used in Google's experiment), the state $|\psi\rangle$ is described by $2^{50} \approx 10^{15}$ complex numbers. Perfectly simulating a quantum circuit on a classical computer is thus an intractable problem, so we assume that the classical algorithm has resources that are \textit{polynomial} in $n$, rather than exponential in $n$. In other words, the sample $S_{\text{cl}}$ obtained from the classical computer is drawn from an approximation of the actual quantum circuit that does not scale appropriately as we increase the number of qubits. Intuitively, then, we might expect that the bitstrings obtained from the classical algorithm are somehow ``different" from those obtained from the actual quantum circuit, since we can only crudely approximate the circuit to obtain these bitstrings. A reasonable way to quantify this difference is the following: we first ask, ``how probable is it that I would obtain the sample $S_{\text{cl}}$ if I had a perfect representation of the circuit's output state $|\psi\rangle$?" This probability is found by computing the inner product between $|x^{\text{cl}}\rangle$ and $|\psi\rangle$, where $|\psi\rangle$ represents the ``perfect" output state (that is, the output state obtained from the experimental implementation of a perfect, error-free quantum computer). We can then compare Pr($S_{\text{cl}}$) to the probability of obtaining the set $S$ produced by the quantum computer Pr($S$). If we give our classical algorithm more resources and/or increase the error rate in our quantum circuit (so our output state $|\psi\rangle$ isn't ``perfect"), we should see these probabilities converge to one another, so Pr$(S) \approx$ Pr$(S_{\text{cl}})$.

If, however, we can make the per-gate error rate of our quantum circuit ``low enough," we observe a difference between these probabilities! For now assuming a quantum circuit with no error, we can quantify this difference as follows: 
    \begin{equation}
        \mathbb{E}_U \left [ \log{ \left ( \frac{\text{Pr}(S)}{\text{Pr}(S_{\text{cl}})} \right )} \right ] \approx m
        \label{logPr}
    \end{equation}
where $m = |S| = |S_{\text{cl}}|$ is the number of sampled bitstrings, and $\mathbb{E}_U$ is read ``the expectation value taken over an ensemble of random unitary operators" (we'll explain more about this in the next section). This is one of the key results from \cite{NEARARXIV}, and a good portion of this article is spent deriving this equation. 

Why is this equation important? Here's what it's telling us: \textit{if your quantum circuit is good enough--that is, the per gate error rate is low--and you perform the sampling experiments described above for a large set of random unitary matrices $\{ U \}$, you'll find (on average) that} $\log{\text{Pr}(S)} - \log{\text{Pr}(S_{\text{cl}})} \approx m$. In other words, this equation provides a check that our quantum computer is actually working by quantifying the difference we \textit{should} see between the quantum circuit and classical simulation. It's sort of a prerequisite for quantum supremacy: if we want to achieve quantum supremacy via the metrics we'll derive in this article, we should have a quantum circuit with a low enough error rate such that this equation approximately holds. 

With all of this in mind, we can set up a roadmap for the next few sections: 
    \begin{enumerate}
        \item We'll first discuss the task of sampling random unitary matrices, and describe why this task is difficult and important for our quantum supremacy experiments. 
        \item Then, we'll show how uniformly sampling random matrices leads to something called the \textit{Porter-Thomas distribution}, which describes how the probabilities $p_{U}(x)$ are distributed if we consider sampling over many random unitary matrices. 
        \item After deriving the Porter-Thomas distribution, we'll be in a position to derive equation \ref{logPr}.  
    \end{enumerate}
After deriving \ref{logPr}, we'll be able to precisely define when quantum supremacy is achieved.

\section{Random matrix theory and the Porter-Thomas distribution}

In our quest to derive equation \ref{logPr}, we have to start at the bottom by defining the task of sampling random unitary operators to define a random quantum circuit. Recall that a unitary operator acts on states in a Hilbert space, and has the property that $U^{\dagger}U = \mathbb{I}$, so inner products between states are preserved. In quantum mechanics, we use matrix representations of unitary operators to act on state vectors and evolve them in time. For dynamical quantum systems, the unitary evolution operator is derived from the time-dependent Schrödinger equation and can be written in terms of the system's Hamiltonian. 

When considering quantum circuits, however, we can interpret the evolution operator as consisting of gate operations on our input qubits. We mentioned this briefly in the previous section, but I think it's worth looking at a simple example just to get an idea of how this process works. Consider the simple circuit shown below: 
    $$\Qcircuit @C=1em @R=.7em {
      & \lstick{\ket{0}} & \gate{H} & \gate{X} & \rstick{\ket{+}} \qw \\
      & \lstick{\ket{0}} & \gate{H} & \gate{X} & \rstick{\ket{+}} \qw 
    }$$
In this case, our evolution operator is given by $U = XH \otimes XH$, which is a $4 \times 4$ matrix acting on the input state $|0\rangle \otimes |0\rangle = (1, 0, 0, 0)^T$. In general, for a system consisting of $n$ qubits, our evolution operator can be represented by a $2^n \times 2^n$ unitary matrix. How is this formalism useful when considering a \textit{random} quantum circuit? It's first helpful to define exactly what ``random" means: in this context, we start with some fixed input state (generally $|\psi_0\rangle = \bigotimes_{i = 1}^{n} |0\rangle$) and apply randomly selected gates to the qubit system. For example, beginning with the state $|\psi_0\rangle$, we could uniformly sample from the universal quantum gate set\footnote{$H$ is the Hadamard gate, $P$ is the phase gate, $CNOT$ is the controlled-NOT gate, and $T$ is the $\pi/8$ gate. See \cite{MIKEIKE} for more information on universal gate sets.} $\{ H, P, CNOT, T, \mathbb{I} \}$ to obtain a gate to apply to qubit $i$ (or qubits $i$ and $i + 1$ if $CNOT$ is selected), and then sample again for qubit $i + 1$ until $i = 2^n - 1$. After repeating this process for a certain number of cycles, we'll have constructed a random unitary operator $U$ that can evolve our input qubit state $|\psi_0\rangle$.

Sampling from a universal gate set is a rather ``physics-based" way to obtain a random quantum circuit, and we can also analyze this problem from the mathematical framework of \textit{random matrix theory}. It was developed in the mid-20th century to help predict the energy spectra of heavy nuclei, and since then has found many applications in quantum chaos and the study of dynamical quantum systems \cite{EIGEN}. Although we don't have the space here to complete a detailed review of random matrix theory, some of the basic ideas are fundamental to the quantum supremacy experiment. To begin, let's consider the space of all $2^n \times 2^n$ complex-valued matrices $\mathbb{C}^{2^n \times 2^n}$; we can think of this as a $2\cdot 2^{2n}$ dimensional Euclidean space, where a point $(x_1, x_2, ..., x_M)$ for $M = 2^{2n + 1}$ and $x_i \in \mathbb{R}$ completely defines a complex matrix \cite{RMTREVIEW}. For quantum computing we're constrained to unitary matrices, so we can consider the subspace $\mathbb{U}^{2^n \times 2^n} \subset \mathbb{C}^{2^n \times 2^n}$ that consists of all \textit{unitary} $2^n \times 2^n$ matrices. Constructing a random quantum circuit is equivalent to sampling a unitary matrix from $\mathbb{U}^{2^n \times 2^n}$, which we can do by randomly selecting a point $(x_1, x_2, ..., x_M) \in \mathbb{U}^{2^n \times 2^n}$. 

So far, so good. There's a key constraint here that wasn't made clear in the last paragraph, but is critical for the experiment: we must sample points $(x_1, x_2, ..., x_M) \in \mathbb{U}^{2^n \times 2^n}$ from a \textit{uniform} distribution over the space of unitary matrices $\mathbb{U}^{2^n \times 2^n}$. This just encodes the idea that we shouldn't have a preference for any given unitary matrix, and is equivalent to our requirement above that we \textit{uniformly} sample from the universal quantum gate set when using the ``physics-based" method to construct a random quantum circuit \cite{HAAR}. How do we ensure that our unitary matrices are sampled uniformly? We have to define an appropriate \textit{measure} on the space $\mathbb{U}^{2^n \times 2^n}$. As a disclaimer, I am by no means an expert on measure theory, so I'll here only discuss the basic conceptual ideas behind using measures to sample matrices. In fact, we'll largely follow the paper \cite{DYSON}, in which Freeman Dyson introduced a framework for defining uniform sampling over the space of unitary matrices.

The overall idea is that we need some way to define ``sampling from a uniform distribution over unitary matrices." This is highly non-trivial for an arbitrary unitary matrix, but is actually straightforward for the set $U(1)$ of $1 \times 1$ unitary matrices, so we'll use this as a toy system\footnote{Inspiration for this toy system came from a discussion on StackExchange: \url{https://math.stackexchange.com/questions/494225/what-is-haar-measure}.} to get some intuition first. The set $U(1)$ just represents all complex numbers $u \in U(1)$ such that $u^{\dagger}u = \mathbb{I}_{1 \times 1} = 1$. Since the adjoint operator is defined as the conjugate transpose when we're representing unitary operators by matrices, and the transpose of a scalar just gives the scalar back, this reduces to the condition $u^*u = 1 \rightarrow |u|^2 = 1$. In other words, $U(1)$ just defines the unit circle in the complex plane, and its elements can be parameterized by an angle $\phi$ such that $u(\phi) = e^{i\phi}$. If we want to sample ``matrices" from a uniform distribution over $U(1)$, it's reasonable to construct this uniformity as follows: \textit{if I have an interval $\Delta \phi$ on the unit circle, I should have an equal probability of selecting a $1 \times 1$ matrix $u$ from $\Delta \phi$ as I do from $\Delta \phi + \phi_0$ for any $\phi_0 \in [0, 2\pi]$.} In other words, no element of $U(1)$ can be selected with a higher probability than any other element. I'm now going to define an integration measure $\mu = d\phi$ on this unit circle, where $d\phi$ represents an infinitesimal angular interval. The probability $\mathcal{P}$ that an element $u \in U(1)$ belongs to the infinitesimal interval $d\phi$ is given by:
    $$ \mathcal{P}(u \in d\phi) = \frac{1}{V} \cdot \mu = \frac{1}{V} \cdot d\phi $$
where $V$ is the ``volume" of the space $U(1)$, which in this case is trivial: 
    $$ V = \int \mu = \int_{0}^{2\pi} d\phi = 2\pi $$
We call the quantity $\eta(u) \equiv d\phi/2\pi$ the \textit{Haar measure} on the set of $1 \times 1$ unitary matrices $U(1)$. Let's analyze a few important properties of the Haar measure; we first note that
    $$ \int \mathcal{P}(u \in d\phi) = \int \eta(u)  = \int_{0}^{2\pi} \frac{d\phi}{2\pi} = 1 $$
This is just telling us that our chosen $u \in U(1)$ must exist \textit{somewhere} on the unit circle, or $\mathcal{P}(u \in U(1)) = 1$. Next, suppose we have an element $u = e^{i\phi}$ that we shift via multiplication by a constant element $u_0 = e^{i\phi_0}$, so $u_0 u = e^{i\phi_0}e^{i\phi} = e^{i(\phi + \phi_0)}$. We can compute the effect of the shift $\phi \rightarrow \phi + \phi_0$ on our Haar measure: 
    $$ \eta(u_0 u) = \frac{d(\phi + \phi_0)}{2\pi} = \frac{d\phi}{2\pi} = \eta(u) $$
So the measure is invariant under transformation by a $1 \times 1$ unitary matrix, which just expresses the fact that $\mathcal{P}(u \in d(\phi + \phi_0)) = \mathcal{P}(u \in d\phi)$; this is exactly how we defined uniformity above! It turns out that the Haar measure $\eta$ can be generalized to many different types of spaces in mathematics, and is defined by the conditions 
    $$ \int \eta(g) = 1  $$
and $\eta(g_0 g) = \eta(g)$, where $g$ represents a member of some space over which the Haar measure is defined (in math, it is often over some topological group). In particular, we can generalize the above discussion to the subspace $\mathbb{U}^{2^n \times 2^n} \subset \mathbb{C}^{2^n \times 2^n}$ of unitary $2^n \times 2^n$ matrices by considering $g = U$ to be elements of $\mathbb{U}^{2^n \times 2^n}$. Note that the condition $\eta(U_0U) = \eta(U)$ ensures uniformity, so no unitary matrix is ``more likely" than any other matrix. Although it's very difficult to parameterize the Haar measure for a general $2^n \times 2^n$ matrix (it can, however, be done, see for example \cite{GENERATE} or \cite{KAROL}), the general structure is similar to what we wrote down above: 
    $$ \eta(U) = \frac{\mu}{V} $$
where $\mu$ is some integration measure over $\mathbb{U}^{2^n \times 2^n}$ and $V$ is the volume of this space. 

Why did we spend so much time talking about the Haar measure? The condition of uniformity is \textit{incredibly important} when we're sampling unitary matrices, and the Haar measure is how we \textit{define} uniformity over a space of matrices. Without this measure, the phrase ``sampling unitary matrices uniformly from $\mathbb{U}^{2^n \times 2^n}$" has no meaning! In the next section, we'll spend some time analyzing precisely why this uniformity is so important.

\subsection{Uniformity leads to Porter-Thomas}

Let's take a step back and recap what we've accomplished so far: we've defined the task of constructing random quantum circuits as (1) uniformly sampling gates from a universal gate set, or (2) uniformly sampling unitary matrices from the space $\mathbb{U}^{2^n \times 2^n}$. We've seen that in order to define uniformity over a space of matrices, we have to construct an appropriate \textit{Haar measure} that is invariant under unitary transformations. Now that we know how to define uniformity, it's logical to ask why this uniformity is useful. In some sense, uniformity is useful because it's what we mean (colloquially, at least) when we say ``random." If I'm drawing samples randomly from some set, I intuitively take this to mean ``no element in the set has a higher probability of being drawn than any other element." So it's logical to conclude that constructing a random quantum circuit means, ``no quantum circuit has a higher probability of being constructed than any other circuit." Although this is true, it doesn't quite answer the question of why we want a random quantum circuit in the first place. We could imagine trying to design a similar experiment where the quantum circuit was fixed, or where the unitary matrix was drawn from some other distribution (for example, a Gaussian-like distribution) over the space of unitary matrices. It turns out that the condition of uniformity is necessary in order to derive equation \ref{logPr}, and ultimately allows us to show a gap in the performance of the quantum computer and classical computer. 

This power of the uniform distribution over unitary matrices is rooted in something called the \textit{Porter-Thomas distribution}. We mentioned above that after sampling a bitstring $x$ from our random quantum circuit, we can compute the probability of obtaining this bitstring as $p(x) = |\langle x | \psi \rangle |^2$. Clearly $x$ in this case is a discrete random variable, and the associated sample space is the set of bitstrings of length $n$. What's interesting is that we can also consider $p(x)$ to be a continuous random variable, if we consider sampling over multiple random quantum circuits. As an example, suppose that I have a set of $\ell$ quantum circuits defined by the unitary matrices $U_1, U_2, ..., U_{\ell}$, where each $U_i$ was sampled from the Haar measure over $\mathbb{U}^{2^n \times 2^n}$. For a given bitstring $x$, I can compute $p_i(x) \equiv p_{U_i}(x)$ for each quantum circuit $U_i$. Can we estimate the \textit{expected value} of $p(x)$ over this set of $\ell$ sampled unitary matrices?

``Well, Sean," you might say, ``let's suppose that we have 50 qubits in our quantum circuit. Then there are $2^{50} \approx 10^{15}$ possible bitstrings; if we're sampling unitary matrices randomly, there's no reason to favor any one bitstring over another. So it's reasonable to expect that $\mathbb{E}[p(x)] \approx 2^{-50}$ for any bitstring $x$, where the expected value is taken over many randomly sampled unitary matrices. In this case, we could treat $p(x)$ as a continuous random variable drawn from the distribution $\delta(p - 2^{-50})$, while the bitstrings can be considered discrete random variables drawn from a uniform distribution." This is excellent logic, and it turns out that if we consider a set of randomly selected \textit{classical} functions $f$ mapping Boolean inputs to $\{ 0, 1 \}^n$, it's exactly correct. No output bitstring is favored, so $\mathbb{E}[p(x)] = 2^{-n}$ for any bitstring $x$.

What's amazing is the following fact: \textit{in the quantum case, the expected value of $p(x)$ is \textbf{not} equal to $2^{-n}$, and the bitstrings are \textbf{not} drawn from a uniform distribution.} As long as our circuit can operate with a sufficiently low error rate, and we're drawing quantum circuits uniformly, the $p(x)$ are distributed according to something called the Porter-Thomas distribution. Letting $\mathcal{P}(p)$ denote the probability density function over the continuous variable $p = p(x) = |\langle x | \psi \rangle |^2$ for some state $|\psi \rangle$ obtained from a random quantum circuit, we have 
    $$ \mathcal{P}(p) = N e^{-Np} $$
where $N = 2^n$ is the dimensionality of the Hilbert space. Note in particular that 
    $$ \mathbb{E}[p(x)] = \int dp \ p Ne^{-Np} \neq 2^{-n} $$
This distribution is critical to deriving equation \ref{logPr}, and it can only be derived if we assume a truly random quantum circuit, where unitary matrices are sampled from the Haar measure. We'll spend the next section deriving the Porter-Thomas distribution, and then finally show how it can be used to derive equation \ref{logPr}. 

\subsection{Deriving Porter-Thomas}

To derive Porter-Thomas, I'll largely be following the derivation in \cite{SUPP}, but will provide some extra details on the calculations for those (like me) who are rusty in their complex analysis. If you want a refresher before diving in, I'd recommend the lightning-fast review in \cite{AMATEUR}. Pay special attention to the Cauchy residue theorem and Cauchy's integral formula, as we'll be using these to prove our main result! 

Suppose we have an $n$-qubit system, and we choose a wavefunction $|\psi\rangle$ at random in the $2^n$ dimensional Hilbert space representing this system. Note that this is equivalent to sampling a unitary $U$ from the Haar measure and applying it to some initial state $|\psi_0\rangle$; we've encoded uniformity in the assumption that our wavefunction is chosen \textit{at random} from our Hilbert space. In the computational basis, we can write this random wavefunction as: 
    $$ |\psi\rangle = \sum_{x} (a_x + ib_x) |x\rangle $$
where $x$ labels the orthonormal basis states for the Hilbert space, and $a_x, b_x \in \mathbb{R}$. We also must enforce the normalization condition: 
    $$ \sum_{x} a_{x}^2 + b_{x}^2 = 1 $$
If we want to calculate $p(x_0)$ for some basis state $|x_0\rangle$ of $|\psi\rangle$, we simply calculate $p(x_0) = |\langle x_0 | \psi \rangle |^2 = a_{x_0}^2 + b_{x_0}^2$. How do we calculate the probability $\mathcal{P}(p)$ of obtaining $p(x_0)$? One way to do this is to ``count" the number of normalized states with $p(x_0) = a_{x_0}^2 + b_{x_0}^2$, and divide that by the total number of normalized states in our Hilbert space. Of course, this doesn't quite work since there are an infinite number of states that satisfy both conditions, so let's modify this slightly. Instead of talking about individual states, let's talk about the \textit{volume} of Hilbert space with (1) states satisfying $p(x_0) = a_{x_0}^2 + b_{x_0}^2$ and the normalization condition, and (2) states only satisfying the normalization condition. In other words, the space of normalized states with $p(x) = a_{x_0}^2 + b_{x_0}^2$ constitutes a subspace of our overall Hilbert space; computing the volume of this subspace, and dividing by the volume of the subspace with only normalized states, should give us an estimate of $\mathcal{P}(p)$. So we have: 
    $$ \mathcal{P}(p) = \frac{\text{Vol}(\mathcal{H}_{p, 1})}{\text{Vol}(\mathcal{H}_1)} $$
where $\mathcal{H}_{p, 1}$ is the subspace with normalized states of probability $p$, and $\mathcal{H}_{1}$ is the subspace of normalized states. To compute these volumes, we use delta functions to ``pick out" the relevant states while integrating over $a_x$ and $b_x$: 
    \begin{align*}
        &\text{Vol}(\mathcal{H}_{p, 1}) = \int_{-\infty}^{\infty} \prod_x da_x db_x \ \delta \left ( \sum_{x} a_{x}^2 + b_{x}^2 - 1 \right ) \delta(a_{x_0}^2 + b_{x_0}^2 - p) \\
        &\text{Vol}(\mathcal{H}_{1}) = \int_{-\infty}^{\infty} \prod_x da_x db_x \ \delta \left ( \sum_{x} a_{x}^2 + b_{x}^2 - 1 \right )
    \end{align*}
The first delta function enforces the normalization condition, while the second delta function in $\text{Vol}(\mathcal{H}_{p,1})$ picks out states $|x\rangle$ such that $p(x) = p$. If we can evaluate these integrals, we can obtain $\mathcal{P}(p)$, which turns out to be the Porter-Thomas distribution! 

Let's warm up with a calculation of the denominator $\text{Vol}(\mathcal{H}_1)$ of $\mathcal{P}(p)$. First note that we can write the delta function in the integrand as an integral: 
    $$ \delta \left ( \sum_{x} a_{x}^2 + b_{x}^2 - 1 \right ) = \frac{1}{2\pi} \int_{-\infty}^{\infty} dt \ e^{it( \sum_{x} a_{x}^2 + b_{x}^2 - 1)} $$
This is a standard result that can be found in most analysis and mathematical physics textbooks. We can expand the sum in the exponential to obtain a product: 
    $$ \delta \left ( \sum_{x} a_{x}^2 + b_{x}^2 - 1 \right ) = \frac{1}{2\pi} \int_{-\infty}^{\infty} dt \ e^{-it} \prod_{x} e^{it(a_{x}^2 + b_{x}^2)} $$
Let's plug this equation back into the denominator of $\mathcal{P}(p)$, so we have: 
    \begin{align*}
        \text{Vol}(\mathcal{H}_1) &= \int_{-\infty}^{\infty} \prod_{x} da_x db_x \ \left ( \frac{1}{2\pi} \int_{-\infty}^{\infty} dt \ e^{-it} \prod_{x} e^{it(a_{x}^2 + b_{x}^2)} \right ) \\
        &= \frac{1}{2\pi} \int_{-\infty}^{\infty} dt \ e^{-it} \left ( \int_{-\infty}^{\infty} \prod_{x} da_x db_x \ e^{it(a_{x}^2 + b_{x}^2)} \right )
    \end{align*}
Let's expand the term in parentheses; notice that this can be written as: 
    \begin{align}
        \left ( \int_{-\infty}^{\infty} \prod_{x} da_x db_x \ e^{it(a_{x}^2 + b_{x}^2)} \right ) = \prod_{x} \int_{-\infty}^{\infty} da_x \ e^{ita_{x}^2} \cdot \int_{-\infty}^{\infty} db_x \ e^{itb_{x}^2}
    \end{align}
The integrals are each a standard Gaussian integral with a value of $\sqrt{i\pi/t}$; since we have a product over $N = 2^n$ possible bitstrings, this gives: 
    $$ \prod_{x} \int_{-\infty}^{\infty} da_x \ e^{ita_{x}^2} \cdot \int_{-\infty}^{\infty} db_x \ e^{itb_{x}^2} = \left ( \frac{i\pi}{t} \right )^N $$
Great! Now we can finally plug this back into our equation for the denominator above, which gives the following integral: 
    $$ \text{Vol}(\mathcal{H}_1) = \frac{1}{2\pi} \int_{-\infty}^{\infty} dt \ e^{-it} \left ( \frac{i\pi}{t} \right )^N = \frac{(i\pi)^N}{2\pi} \int_{-\infty}^{\infty} dt \ \frac{e^{-it}}{t^N} $$
Okay, now's the fun part: we have to evaluate the integral over $t$. It turns out that we can compute this integral by defining a corresponding \textit{contour integral} in the complex plane, and then use \textit{Cauchy's residue theorem} to very easily evaluate the contour integral. As long as the contour integral is equal to the integral we ultimately want to solve (the one over $t$), we don't actually have to do much work! The trick, unfortunately, is finding a useful contour integral that does correspond to $\text{Vol}(\mathcal{H}_1)$ above. The supplementary material \cite{SUPP} completely skips this part of the calculation, so I'm going to walk through it here for completeness. If you're like me and have never taken a formal complex analysis course, I'd recommend brushing up on the Cauchy residue theorem before continuing; there are plenty of excellent textbooks and lecture notes that cover the subject (I used \cite{656} while refreshing myself). The good news is that the theorem itself is fairly simple, but actually applying it to our friendly $\text{Vol}(\mathcal{H}_1)$ integral takes a little bit more thought. 

Okay, let's do this. The first step is to define a corresponding contour integral in the complex plane, and a reasonable choice might seem to be: 
    $$ \int_{-\infty}^{\infty} dt \ \frac{e^{-it}}{t^N} \rightarrow \oint_{\mathcal{C}'} dz \ \frac{e^{-iz}}{z^N} $$
In this case, $\mathcal{C}'$ is the contour we'll use to evaluate the complex integral: along the real axis to $\pm \infty$, with the ``ends" of the real line connected via a half-circle in the lower complex plane. Unfortunately, this choice \textit{doesn't work} because the complex function we're integrating doesn't have a \textit{simple} pole at $z = 0$. You can read about the details of this in your favorite complex analysis textbook; the short story is that our contour $\mathcal{C}'$ can't pass through the pole at $z = 0$, but our contour has to travel over the real axis because we ultimately want to integrate over all real numbers. Usually, this is fixed by taking a tiny detour (in the form of a half-circle of radius $\epsilon$) around the pole at $z = 0$, and then showing that the integral around this half-circle vanishes as $\epsilon \rightarrow 0$. If you try this on the complex integral above, you can show that this ``half-circle integral" doesn't vanish in the $\epsilon \rightarrow 0$ limit. In fact, the integral depends on a factor of $1/\epsilon^{N - 1}$, so it actually blows up!  

\begin{figure}
    \centering
    \includegraphics[scale=0.55]{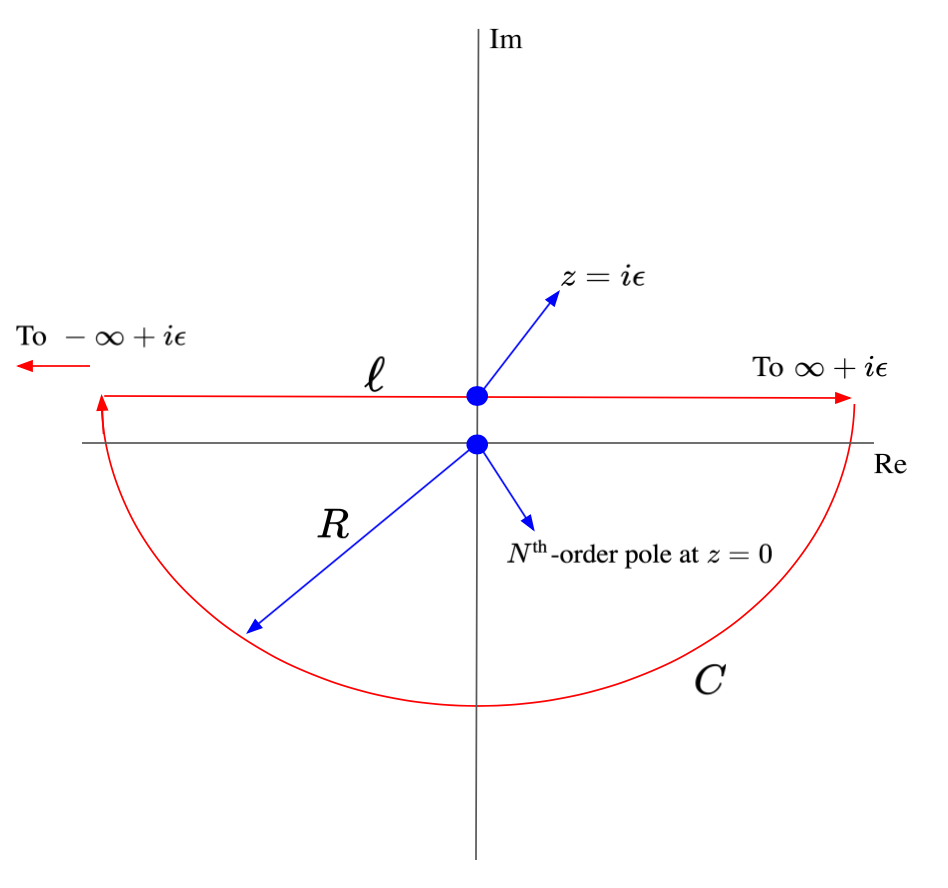}
    \caption{The contour $\mathcal{C}$ used for evaluating the integral of $e^{-it}/t^N$. Note that $\mathcal{C} = \ell + C$, and that $R \rightarrow \infty$ as $\ell$ is extended to $\pm \infty + i\epsilon$.}
    \label{contour}
\end{figure}

To fix this, let's add a small, constant, imaginary offset to the bounds of integration, noting that 
    $$ \int_{-\infty}^{\infty} dt \ \frac{e^{-it}}{t^N} = \lim_{\epsilon \rightarrow 0} \int_{-\infty + i\epsilon}^{\infty + i\epsilon} dt \ \frac{e^{-it}}{t^N}  $$
This shifts our integration contour to just above the real axis, so it no longer passes through an $N$th order pole at $z = 0$. If we set up the contour $\mathcal{C}$ given in figure \ref{contour}, we can use Cauchy's ``integral formula for derivatives" to obtain, integrating clockwise: 
    $$ \oint_{\mathcal{C}} dz \  \frac{e^{-iz}}{z^N} = \frac{-2\pi i}{(N - 1)!} \cdot \frac{d^{N - 1}}{dz^{N - 1}} \left [ e^{-iz} \right ]_{z = 0} = \frac{-2\pi i}{(N - 1)!} \cdot (-i)^{N - 1} $$
Before continuing, let's briefly verify that the contour integral we evaluated is ultimately equal to the original integral we wanted to compute. We can split the contour into a linear part $\ell$ and a semicircular part $C$:
    $$ \oint_{\mathcal{C}} dz \  \frac{e^{-iz}}{z^N} = \int_{\ell} dz \ \frac{e^{-iz}}{z^N} + \int_{C} dz \ \frac{e^{-iz}}{z^N} $$
The integral over $\ell$ is obviously equal to our original integral, since we're integrating over all real numbers with an added constant offset $i\epsilon$, for an arbitrarily small $\epsilon$. We thus have to show that the integral over the semicircle goes to zero as we extend our linear integration limits to $(\pm \infty) + i\epsilon$. To do this, we'll use \textit{Jordan's lemma}, which you can find in any textbook or lecture notes on complex analysis (for example, \cite{656}). Note that our semicircle's radius $R \rightarrow \infty$ as we extend the linear integration limits, and that $1/z^N \rightarrow 0$ for $z \rightarrow \infty$. Jordan's lemma thus tells us that the integral over $C$ approaches 0, as long as $C$ is in the lower half of the complex plane. This verifies the desired equality: 
    $$ \oint_{\mathcal{C}} dz \  \frac{e^{-iz}}{z^N} = \int_{-\infty + i\epsilon}^{\infty + i\epsilon} dt \ \frac{e^{-it}}{t^N} $$
Again note that we can choose $\epsilon$ to be as small as we want, so the integral above ultimately converges to our original integral without an $i\epsilon$ offset in the integration bounds. 

\begin{figure}
    \centering
    \includegraphics[scale=0.5]{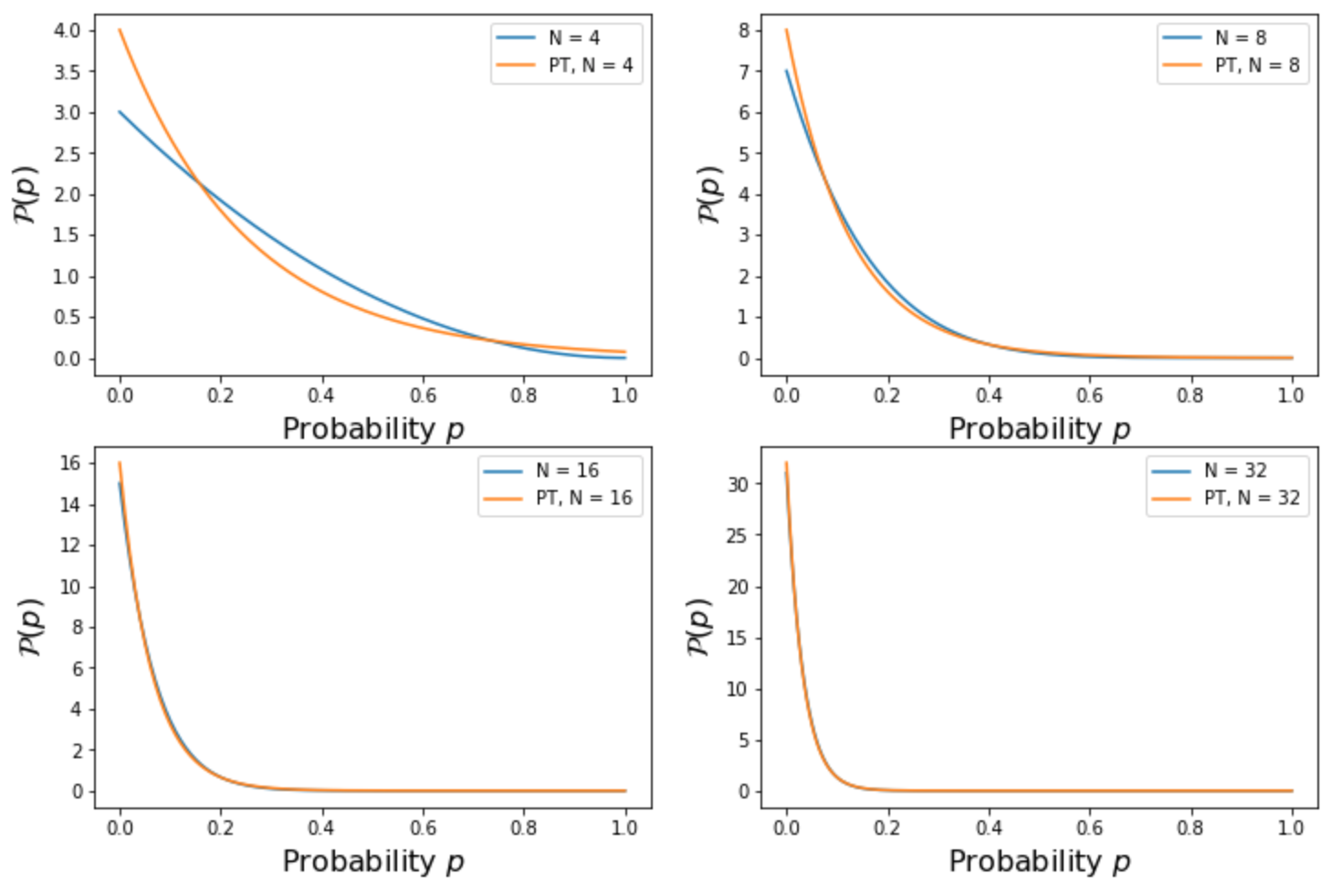}
    \caption{Convergence of $\mathcal{P}(p)$ (in blue) to the Porter-Thomas distribution (in orange) as $N = 2^n$ increases. Note that we already see excellent agreement for only 5 qubits!}
    \label{PT}
\end{figure}

With all of the verification out of the way, we can plug this result back into our equation for $\text{Vol}(\mathcal{H}_1)$ to give: 
    $$ \text{Vol}(\mathcal{H}_1) = \frac{(i\pi)^N}{2\pi} \cdot \frac{-2 \pi i}{(N - 1)!} \cdot (-i)^{N - 1} = \frac{\pi^N}{(N - 1)!} $$
This is exactly what's given in \cite{SUPP}! The integration of the numerator is almost exactly identical; the only complication is the inclusion of the $\delta(a_{x_0}^2 + b_{x_0}^2 - p)$ term, which introduces an integral over another dummy variable (similar to the integral over $t$ above). Since the math is essentially identical, I'm not going to go through it here, as you have all the tools you need to complete it on your own! Regardless, we'd find that the numerator is given by: 
    $$ \text{Vol}(\mathcal{H}_{p, 1}) = \frac{\pi^N}{(N - 2)!} (1 - p)^{N - 2} $$
Putting it all together, we have: 
    $$ \mathcal{P}(p) = \frac{(N - 1)!}{(N - 2)!} (1 - p)^{N - 2} = (N - 1)(1 - p)^{N - 2} $$
We're almost there! The authors claim that for large $N$, this distribution approaches the Porter-Thomas form we mentioned earlier. Instead of verifying this analytically, I decided to verify it numerically to visualize the convergence of the two functions as $N = 2^n$ grows large. The results are shown in figure \ref{PT}; note that there is excellent agreement between the two functions when $N = 32$, corresponding to only 5 qubits! So we can confidently say that for $N$ ``large" ($\gtrsim 32$), we have
    $$ \mathcal{P}(p) \rightarrow Ne^{-Np} $$
exactly as claimed in \cite{SUPP}. Great! \textbf{We now understand why sampling random unitary matrices from the Haar measure ultimately gives rise to bitstring probabilities that are distributed according to the Porter-Thomas distribution.} I want to stop here and acknowledge what we've done: the \textit{entire} quantum supremacy experiment ultimately rests on this theoretical foundation! We depend on a difference in bitstring probabilities between the quantum and classical cases to ultimately show that quantum supremacy has been achieved, so this is an extremely powerful result. As a next step, we're going to see exactly \textit{how} we can use the Porter-Thomas distribution to derive equation \ref{logPr}. 

\section{Demonstrating quantum supremacy}

As we noted in our Roadmap above, equation \ref{logPr} helps us define the task of quantum supremacy: if our quantum computer has a low enough error rate, and we are effectively sampling unitary matrices from the Haar measure, we should find that
    $$ \mathbb{E}_U \left [ \log{ \left ( \frac{\text{Pr}(S)}{\text{Pr}(S_{\text{cl}})} \right )} \right ] \approx m $$
where $m$ is the total number of bitstrings sampled. To derive this equation, we'll be following section II of \cite{NEARARXIV}, so you should skim this part of the paper if you haven't already. 

To begin the derivation, we're going to use something called the \textit{asymptotic equipartition property}. We won't rigorously derive this property here, although we'll discuss the derivation conceptually below. The mathematically inclined reader will find a more rigorous proof in the excellent lecture notes \cite{376A}, from the class EE376A at Stanford University. From these notes, we use the following definition: 

\textbf{Definition 1.} \textit{For any $\epsilon > 0$, a set of bitstrings $S$ of size $m$ (where the bitstrings are sampled from a random quantum circuit) is \textbf{epsilon-typical} if the following holds:} 
    $$ \left | -\frac{1}{m}\log{\text{Pr}(S)} - H(p(x))  \right | \leq \epsilon $$
\textit{where $H(p(x))$ is the information entropy, defined by:} 
        $$ H(p(x)) = -\sum_{i = 1}^{N} p(x_i) \log{p(x_i)}  $$
\textit{Here $N = 2^n$ is the total number of possible bitstrings.} 

In order to use this property in our derivation, we'll have to use the following theorem: 

\textbf{Theorem 1.} \textit{Let $\mathcal{E}_{m, \epsilon}$ denote the set of all $S$ (with $|S| = m$) that are epsilon-typical. Then for all $\epsilon > 0$, we have the following:} 
    $$ P(S \in \mathcal{E}_{m, \epsilon}) \rightarrow 1 \ \text{as} \ m \rightarrow \infty $$
\textit{Here we use $P(S \in \mathcal{E}_{m, \epsilon})$ to mean ``the probability that $S$ is contained in the epsilon-typical set," to distinguish from the probability distribution over bitstring sets} $\text{Pr}(S)$.

Together, definition 1 and theorem 1 define the asymptotic equipartition property. Why is this useful? It tells us that if the number of sampled bitstrings $m$ is large enough, we can effectively set $\log{\text{Pr}(S)} = -mH(p) - m\epsilon$. Since we can choose $\epsilon$ to be arbitrarily small, we'll make the approximation 
    $$ \log{\text{Pr}(S)} \approx -mH(p) $$
to make the following derivation easier. In other words, we'll assume $\epsilon \rightarrow 0$ faster than $m \rightarrow \infty$. Our next step is to evaluate $H(p)$, so we can obtain an explicit expression for $\log{\text{Pr}(S)}$. Ultimately we'll want to obtain another explicit expression for $\log{\text{Pr}(S_{\text{cl}}})$, and then we'll take the expectation value over all unitary matrices sampled from the Haar measure to prove equation \ref{logPr}.

To evaluate $H(p)$, we have to reference \cite{SUPP}, which contains important information for evaluating this sum given that $p \sim Ne^{-Np}$, as we proved above for large $N$. Suppose we have some function $f(p) = f(p(x))$; by the properties of the Dirac delta function, we must have: 
    $$ \sum_{x \in \{0, 1\}^n} f(p(x)) = \int_{0}^{1} dp' \left ( \ \sum_{x \in \{0, 1\}^n} \delta(p(x) - p') f(p') \right ) $$
Note that the integration is over some dummy variable $p'$, and that we can restrict the integration to the interval $[0,1]$ because $0 \leq p = p(x) \leq 1$. Also note that in our notation, we'll use $p$ and $p(x)$ interchangeably. We'll substitute for the delta function by first noticing the following for large $N$: 
    $$ \frac{1}{N} \int_{a}^{b} dp' \ \left ( \sum_{x \in \{ 0, 1 \}^n} \delta(p(x) - p') \right ) = \frac{1}{N} \int_{a}^{b} dp' \ \left ( \sum_{p} \delta(p - p') \right ) = \int_{a}^{b} dp \ Ne^{-Np} $$
The notation $\sum_{p}$ is read, ``the sum over all $N$ probabilities $p = p(x)$ observed given a set of $N$ bitstrings $\{x\}$ from a random quantum circuit." It turns out that this equality implies the following equation: 
    $$ \frac{1}{N} \int_{0}^{\infty} dp' \ f(p') \sum_{p} \delta(p - p') = \int_{0}^{\infty} dp \ f(p) Ne^{-Np} $$
We won't provide a rigorous proof of these equalities here, but we do give a justification of the equalities in Appendix A and encourage all readers to analyze that discussion after finishing this section. Regardless, this last expression is extremely useful, as we can substitute as follows: 
    \begin{align*}
        \int_{0}^{1} dp' \left ( \ \sum_{p} \delta(p - p') f(p') \right ) &= \int_{0}^{\infty} dp' \left ( \ f(p') \sum_{p} \delta(p - p') \right ) \\
        &= \int_{0}^{\infty} dp \ f(p) N^2 e^{-Np} \\
        &\Rightarrow \sum_{x \in \{0, 1\}^n} f(p(x)) = \sum_{p} f(p) = N \cdot \mathbb{E}[f(p)]
    \end{align*}
In the first line, we extended the upper limit of integration to $\infty$ because this leaves the value of the integral unchanged (since $0 \leq p(x) \leq 1$, all the delta functions are contained in the interval $[0,1]$, so integrating outside this range just gives 0). Also note that the expectation value is taken over the Porter-Thomas distribution. The above equation is immediately useful to us, as for $f(p(x)) = p(x) \log{p(x)}$, we must have: 
    $$ -\sum_{i = 1}^{N} p(x_i) \log{p(x_i)} = -N \cdot \int_{0}^{\infty} dp \ Np e^{-Np} \log{p} $$
You can evaluate the integral with your favorite integration software, but it turns our to be $\log{N} - 1 + \gamma$, where $\gamma$ is Euler's constant. So we've found that: 
    $$ \boxed{\log{\text{Pr}(S)} = -m(\log{N} - 1 + \gamma)} $$
This is a super important result! We'll use it soon. To move forward, we now need to obtain an analogous expression for $\text{Pr}(S_{\text{cl}})$. To do this, we're actually going to go back to the asymptomatic equipartition property and consider the non-classical set $S$ again. I want to rewrite definition 1 by expanding $\text{Pr}(S)$ as follows: 
    $$ \log{\text{Pr}(S)} = \log{\left ( \prod_{x \in S} p(x) \right )} = \sum_{x \in S} \log{p(x)} $$
Let's also rewrite the entropy $H(p)$: 
    $$ H(p) = -\sum_{i = 1}^{N} p(x_i) \log{p(x_i)} = \mathbb{E}_p[-\log{p(x)}] $$
where $\mathbb{E}_p[\cdot]$ is read ``the expectation value of $\cdot$ over the distribution $p$." We can now rewrite the definition of the epsilon-typical set as follows: \textit{$S$ is epsilon-typical if}
    $$ \left | \frac{1}{m} \sum_{x \in S} (-\log{p(x)}) - \mathbb{E}_p[-\log{p(x)}] \right | \leq \epsilon $$
\textit{for any $\epsilon > 0$}. Let's carefully analyze the structure of this equation; the first term is a sample average of the random variable $(-\log{p(x)})$ for $x \sim p(x)$, while the second term is the expected value of this random variable over the distribution $p(x)$. By the weak law of large numbers, as $m \rightarrow \infty$ the sample average should converge to the expected value, so the above equation should hold for large $m$ (the formal proof of theorem 1 uses essentially this same argument).   

I want to summarize the above discussion as follows: we first sample bitstrings $x \sim p(x)$, where $p(x)$ represents a distribution over all possible bitstrings. We then compute $f(x) = -\log{p(x)}$ for each sampled bitstring, noting that $f(x)$ is an \textit{iid} random variable since the bitstrings $x$ are \textit{iid} based on our experimental setup. We then take the average value of $f(x)$ over our set of sampled bitstrings $S$. If we complete this experiment for large $|S| = m$, we should find that the sample average converges to the expected value of $f(x)$, with respect to the distribution over bitstrings $p(x)$. 

That last paragraph may have seemed redundant, but the logic and notation will be very important in what follows. Let's now consider the set of bitstrings $S_{\text{cl}}$ sampled from the classical algorithm; note that the classical algorithm constructs a probability distribution $p_{\text{cl}}(x) = |\langle x | \psi_{\text{cl}} \rangle |^2$, where $|\psi_{\text{cl}}\rangle$ is the approximate wavefunction constructed by the algorithm with resources polynomial in the number of qubits $n$. Also recall that when calculating $\text{Pr}(S_{\text{cl}})$, we use the distribution $p(x)$ produced by the quantum computer, not the distribution from the classical computer. Consider the following definition: 

\textbf{Definition 2.} \textit{For any $\epsilon > 0$, a set of bitstrings} $S_{\text{cl}}$ \textit{of size $m$ (where the bitstrings are sampled from the classical algorithm) is \textbf{cross-epsilon typical} if the following holds:} 
    $$ \left | -\frac{1}{m}\log{\text{Pr}(S_{\text{cl}})} - H(p_{\text{cl}}, p)  \right | \leq \epsilon $$
\textit{where} $H(p_{\text{cl}}, p)$ \textit{is the cross entropy, defined by:} 
        $$ H(p_{\text{cl}}, p) = -\sum_{i = 1}^{N} p_{\text{cl}}(x_i) \log{p(x_i)}  $$
\textit{As a note, ``cross-epsilon typical" is my own term (not standard in literature) to distinguish from definition 1 above.}

To obtain our expression for $\text{Pr}(S_{\text{cl}})$, we'll use the following theorem: 

\textbf{Theorem 2.} \textit{Let $\mathcal{C}_{m, \epsilon}$ denote the set of all} $S_{\text{cl}}$ \textit{that are cross-epsilon typical. Then for all $\epsilon > 0$, we have the following:} 
    $$ P(S_{\text{cl}} \in \mathcal{C}_{m, \epsilon}) \rightarrow 1 \ \text{as} \ m \rightarrow \infty $$
\textit{Here we use $P(S_{\text{cl}} \in \mathcal{C}_{m, \epsilon})$ to mean ``the probability that} $S_{\text{cl}}$ \textit{is contained in the cross-epsilon typical set," to distinguish from the probability distribution over classical bitstring sets} $\text{Pr}(S_{\text{cl}})$.

Why should this theorem be true? You'll notice that definition 2 and theorem 2 are very similar to definition 1 and theorem 1, with the exceptions that (1) we're working with bitstrings drawn from the classical algorithm, and (2) we've introduced this cross-entropy term. It turns out that using the cross entropy is critical for theorem 2 to hold. To see why, note that the cross entropy can be rewritten: 
    $$ H(p_{\text{cl}}, p) = -\sum_{i = 1}^{N} p_{\text{cl}}(x_i) \log{p(x_i)} = \mathbb{E}_{p_{\text{cl}}}[-\log{p(x)}] $$
In other words, the cross entropy of $p_{\text{cl}}$ with respect to $p$ is obtained by taking the expectation value of $(-\log{p(x)})$ over the distribution $p_{\text{cl}}$. We can expand the $\text{Pr}(S_{\text{cl}})$ term as we did before,  allowing us to rewrite the definition of the cross-epsilon typical set: $S_{\text{cl}}$ \textit{is cross-epsilon typical if}
    $$ \left | \frac{1}{m} \sum_{x \in S_{\text{cl}}} (-\log{p(x)}) - \mathbb{E}_{p_{\text{cl}}}[-\log{p(x)}] \right | \leq \epsilon $$
\textit{for any $\epsilon > 0$}. Let's summarize what this equation is telling us: we first sample bitstrings $x \sim p_{\text{cl}}(x)$, where $p_{\text{cl}}(x)$ represents the classically obtained distribution over all bitstrings. We then compute $f(x) = -\log{p(x)}$ for each sampled bitstring, noting that this is the exact same function we used above when analyzing the epsilon-typical set. We then take the average value of $f(x)$ over our set of classically sampled bitstrings $S_{\text{cl}}$. If we complete this experiment for large $|S_{\text{cl}}| = m$, we should find that the sample average converges to the expected value of $f(x)$, \textit{with respect to the distribution over classical bitstrings $p_{\text{cl}}(x)$.}

That last sentence is critical! Since we're sampling bitstrings from the classical distribution, the expectation value has to be taken over the classical distribution and not the ``quantum" distribution $p(x)$. This is what leads to the introduction of the cross entropy term! The last paragraph also provides a sketch of the proof for theorem 2: by the weak law of large numbers, the sample average should converge to the expectation value for large $m$. Armed with these tools, we note that for large $m$, we can effectively set 
    $$ \log{\text{Pr}(S_{\text{cl}})} \approx -mH(p_{\text{cl}}, p) $$
for an arbitrarily small $\epsilon$. Instead of immediately evaluating the cross entropy, we're going to take the expectation value of this quantity over a set of randomly sampled unitary matrices $\{U\}$:
    $$ \mathbb{E}_{U}[\log{\text{Pr}(S_{\text{cl}})}] = -m \cdot \mathbb{E}_{U}[H(p_{\text{cl}}, p)] = -m \cdot \mathbb{E}_{U} \left [ -\sum_{i = 1}^{N} p_{\text{cl}}(x_i) \log{p(x_i)} \right ] $$
To make this equation simpler, we're going to make an assumption that the authors of \cite{NEARARXIV} make very explicit: since the classical algorithm has resources that are \textit{polynomial} in the number of qubits, the output of the classical algorithm is effectively uncorrelated with the output of the random quantum circuit. They give numerical evidence for this assumption in \cite{NEARARXIV} which we don't have time to cover here, so we'll take this assumption as a given for now. How does this assumption help us? Well, when taking an expectation value over the set of random unitary matrices, we're treating $p(x)$ and $p_{\text{cl}}(x)$ as continuous random variables. If the outputs of the classical and quantum algorithms are uncorrelated, we can treat $p_{\text{cl}}$ and $p$ as \textit{independent} random variables, so the above equation simplifies: 
    $$ -m \cdot \mathbb{E}_{U} \left [ -\sum_{i = 1}^{N} p_{\text{cl}}(x_i) \log{p(x_i)} \right ] = m \cdot \sum_{i = 1}^{N} \mathbb{E}_U [p_{\text{cl}}(x_i)] \cdot \mathbb{E}_U [p(x_i)] $$
Since the $p(x)$ are distributed according to the Porter-Thomas distribution, we can easily compute the expectation value: 
    $$ \mathbb{E}_U [p(x_i)] = \int_{0}^{\infty} dp \ Ne^{-Np} \log{p} = -(\log{N} + \gamma) $$
Since the classical algorithm is uncorrelated with the quantum circuit, we should just have $\mathbb{E}_U[p_{\text{cl}}(x_i)] = p_{\text{cl}}(x_i)$. Substituting, we obtain: 
    $$ m \cdot \sum_{i = 1}^{N} \mathbb{E}_U [p_{\text{cl}}(x_i)] \cdot \mathbb{E}_U [p(x_i)] = -m \cdot (\log{N} + \gamma) \cdot \sum_{i = 1}^{N} p_{\text{cl}}(x_i) = -m \cdot (\log{N} + \gamma) $$
Note that all the classical probabilities must sum to one. So we finally have the following: 
    $$ \boxed{\mathbb{E}_U[\log{\text{Pr}(S_{\text{cl}})}] = -m(\log{N} + \gamma)} $$
Great! Now we can combine our two boxed equations, noting that since we derived $\log{\text{Pr}(S)}$ to be a constant with respect to $p(x)$, we have $\mathbb{E}_U[\log{\text{Pr}(S)}] = \log{\text{Pr}(S)}$. So we finally have: 
    \begin{align*}
        \mathbb{E}_U \left [ \log{ \left ( \frac{\text{Pr}(S)}{\text{Pr}(S_{\text{cl}})} \right )} \right ] &= \mathbb{E}_U[ \log{\text{Pr}(S)} - \log{\text{Pr}(S_{\text{cl}})} ] \\
        &= -m \log{N} + m - m\gamma + m\log{N} + m\gamma \\
        &= m
    \end{align*}
This is exactly what we wanted to show! Sweet! Where do we go from here? Our discussion of the cross-entropy actually leads to a way to experimentally test for quantum supremacy. 

\subsection{Cross entropy benchmarking}

Suppose our classical algorithm simply outputs a uniform distribution over all bitstrings, so $p_{\text{cl}}(x_i) = 1/N \ \forall \ i$. We can easily compute the cross entropy of the uniform sampler and the ideal distribution from the circuit $p(x)$, denoting this quantity by $H_0$:
    \begin{align*}
        H_0 &= -\sum_{i = 1}^{N} \frac{1}{N} \log{p(x_i)} \\
        &= -\frac{1}{N} \cdot N \int_{0}^{\infty} dp \ Ne^{-Np} \log{p} \\
        &= \log{N} + \gamma
    \end{align*}
Note that we used the identity $\sum_{x} f(p(x)) = N \cdot \mathbb{E}[f(p)]$, which we derived earlier. This is interesting because it suggests a way to compare some sampling algorithm, which might be classical or quantum, to an ideal, error-free quantum computer performing the sampling task. Our worst-case scenario, a uniform sampler over bitstrings, corresponds to $H_0 = \log{N} + \gamma$, so all we have to do is compute the cross entropy of our algorithm and compare it to this ``baseline" cross-entropy. The authors of \cite{NEARARXIV} thus define what they call the \textit{cross entropy difference} $\Delta H(p_A)$,
    $$ \Delta H(p_A) \equiv H_0 - H(p_A, p) $$
where $p_A$ is the output distribution of the algorithm $A$ we're testing, and $p$ is the ideal output distribution. We can rewrite this as follows: 
    \begin{align*}
        \Delta H(p_A) &= -\sum_{i = 1}^{N} \frac{1}{N} \log{p(x_i)} + \sum_{i = 1}^{N} p_A(x_i) \log{p(x_i)} \\
        &= \sum_{i = 1}^{N} \left ( p_A(x_i) - \frac{1}{N} \right ) \log{p(x_i)}
    \end{align*}
In the worst-case scenario $p_A(x_i) = 1/N$, we have $\Delta H = 0$, while the best-case scenario of $p_A = p$ gives the following: 
    \begin{align*}
        \Delta H(p) &= \sum_{i = 1}^{N} \left ( p(x_i) - \frac{1}{N} \right ) \log{p(x_i)} \\
        &= \sum_{i = 1}^{N} p(x_i) \log{p(x_i)} - \frac{1}{N} \sum_{i = 1}^{N} \log{p(x_i)} \\
        &= -\log{N} + 1 - \gamma + \log{N} + \gamma \\
        &= 1
    \end{align*}
Note that we've already evaluated each of the sums in this expression earlier in the article. So if the cross-entropy difference is unity, we know that our algorithm $A$ is performing at the same level as an ideal quantum computer on the sampling task. This measure of the algorithm's performance is called \textit{cross-entropy benchmarking}, and can be used to (finally!) quantitatively define the quantum supremacy task. 

\textbf{Demonstrating quantum supremacy.} With an experimental quantum computer, sample uniformly from a universal gate set to obtain a random quantum circuit with output distribution $p_{\text{exp}}$. Compute the cross-entropy difference $\Delta H(p_{\text{exp}})$. Repeat this procedure many times, with a different uniformly sampled quantum circuit each time. After many cycles, compute the parameter 
    $$ \alpha = \mathbb{E}_U[\Delta H(p_{\text{exp}})] $$
where the expectation is taken over an ensemble of random unitary matrices. Quantum supremacy is achieved when 
    $$ \boxed{1 \geq \alpha > C} $$
where $C \equiv \mathbb{E}_U[\Delta H(p^*)]$ is the expected cross-entropy difference of the best possible \textit{classical} algorithm performing the sampling task. That is, $p^*$ is the output distribution of the best possible classical algorithm. 

Awesome! After 18 pages, we finally understand everything that goes into this equation. 

\section{Conclusions}

\begin{figure}
    \centering
    \includegraphics[scale=0.55]{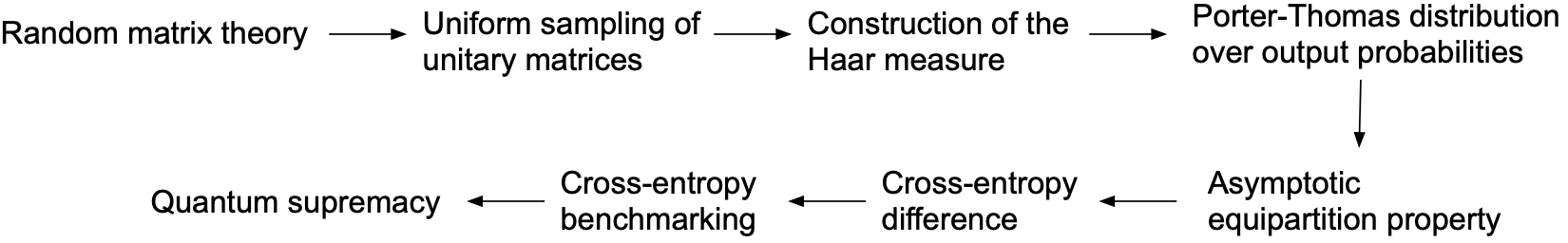}
    \caption{General structure of the topics covered in this article, from random matrix theory to a quantitative definition of quantum supremacy.}
    \label{topics}
\end{figure}

\begin{figure}
    \centering
    \includegraphics[scale=0.55]{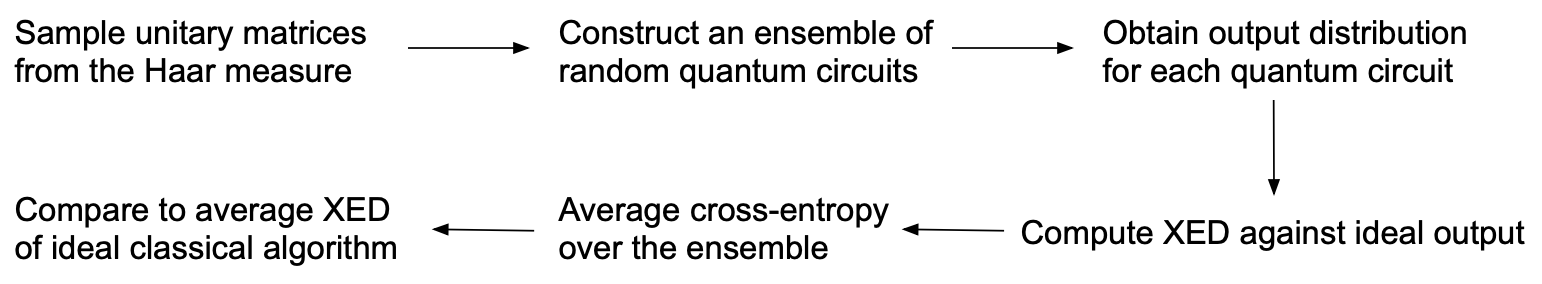}
    \caption{Outline of how to demonstrate quantum supremacy, using the quantitative metric we defined. Note that ``XED" stands for ``cross entropy difference."}
    \label{process}
\end{figure}

To wrap up this article, I want to summarize everything we've studied, since there's quite a long chain of reasoning from random matrix theory all the way to our precise definition of quantum supremacy. I often find it helpful to visually map out how the concepts are related, and have done so in figures \ref{topics} and \ref{process}. We first considered random matrix theory, and thought about how to sample unitary matrices uniformly from the space of $2^n \times 2^n$ unitary matrices. This led to the development of the Haar measure, which defines what ``uniform sampling" over this space means mathematically. With the assumption of uniformly sampled unitaries, we were able to show that the output probabilities of the random quantum circuits are distributed according to the Porter-Thomas distribution. We then derived explicit expressions for the probability distributions over quantum-sampled bitstrings and classically-sampled bitstrings, using the asymptotic equipartition property and the cross-entropy. This naturally led to the definition of cross-entropy benchmarking as a way to compare any algorithm (classical or quantum) to an ideal quantum circuit. Finally, we used cross-entropy benchmarking to precisely define when quantum supremacy is achieved. 

A natural next step is to start analyzing how all of this is carried out experimentally. We unfortunately don't have to space to cover the experimental implementation here; that will have to wait for another article! The implementation is highly non-trivial, and depends very sensitively on \textit{errors} in the quantum circuit. Our entire discussion in this article assumed an error-free quantum computer, an assumption that is certainly not satisfied in experimental quantum computers today. In fact, the authors of \cite{NEARARXIV} note that in the presence of a errors, equation \ref{logPr} depends \textit{exponentially} on the error rate: 
    $$ \mathbb{E}_U \left [ \log{ \left ( \frac{\text{Pr}(S)}{\text{Pr}(S_{\text{cl}})} \right )} \right ] \approx m \cdot e^{-rg} $$
Here $r$ is the per-gate error rate and $g$ is the total number of rates. This is a huge problem, as even a modest increase in $r$ can lead to a drastic reduction in the separation between the output distributions of the quantum and classical cases. A very large portion of \cite{SUPRNAT} is thus dedicated to reducing the error rate so the classical and quantum sampling algorithms can be distinguished. 

We also assumed \textit{perfectly random} quantum circuits in this article, which would require perfect sampling from the Haar measure on $\mathbb{U}^{2^n \times 2^n}$. As noted in \cite{HAAR}, however, constructing a quantum circuit to perform this sampling perfectly requires resources exponential in the number of qubits (that is, the circuit would require $\mathcal{O}(n^2 2^n)$ gates and $\mathcal{O}(2^{2n})$ input parameters). Since the entire theoretical basis for the experiment rests on the fact that we \textit{can't} implement an algorithm with resources exponential in the number of qubits, a perfect sampler from the Haar measure is effectively impossible. An experimental implementation would have to take into account the \textit{pseudo-randomness} of our sampler, and ensure that the limiting distribution of $p(x)$ output from a pseudo-random circuit is still the Porter-Thomas distribution. 

If the restrictions just discussed don't sound bad enough, it gets worse! Let's briefly consider how the quantities $\alpha$ and $C$ would be computed experimentally, in order to actually demonstrate quantum supremacy. In order to compute $\alpha$, we have to somehow obtain the output probability distribution $p_A(x)$ for the quantum circuit $A$ that we're using. We can't obtain this directly from the quantum circuit, as upon measurement the wavefunction $|\psi\rangle$ collapses to give a single state $|x\rangle$ with probability $|\langle x | \psi \rangle|^2$. Measuring thus allows us to \textit{sample} from $p(x)$, but doesn't give us the ``global" view of the distribution we need to compute $\alpha$. One way around this problem is to classically simulate the circuit $A$ and obtain some representation of the output wavefunction $|\psi\rangle$, and then use this to compute the distribution $p_A(x)$ as needed. This is feasible for a circuit with a small number of qubits, but being able to accurately compute $|\psi\rangle$ also implies that $C \approx \alpha$. In other words, if our quantum circuit is small such that we can compute $|\psi\rangle$ accurately with a classical computer, this \textit{also} implies that the classical computer can efficiently simulate the quantum circuit, so we can't possibly be in the quantum supremacy regime where $\alpha > C$. 

A logical solution might be to increase the size of the quantum circuit so that it can no longer be simulated classically, thus guaranteeing that $\alpha > C$. In this case, however, we have no way to obtain the output distribution $p_A(x)$ of our quantum circuit, since we can no longer accurately simulate the output wavefunction $|\psi\rangle$! We have a sort of circular reasoning: to guarantee quantum supremacy we must be in the regime where $\alpha > C$, but in this regime we can't directly compute $\alpha$, so we have no way of proving that we're working in this regime in the first place. The authors of \cite{NEARARXIV} foreshadow their solution to this apparent paradox by claiming: 
    \begin{quote}
        \textit{We argue that the observation of a close correspondence between experiment, numerics and theory would provide a reliable foundation from which to extrapolate} $\alpha$.
    \end{quote}
What does this mean? Unfortunately, I have no idea! As I mentioned in this article's introduction, I wanted to limit my scope here to discussing the experiment's \textit{theoretical} foundations. The experimental basis of the experiment will have to wait for another article. I did, however, want to make sure the difficulty of calculating $\alpha$ was mentioned somewhere, if only to give an appreciation for just how difficult it was to carry out this experiment. 

The Google \textit{et al} team was still able to satisfy and verify the condition for quantum supremacy, which is remarkable given the difficulties we just discussed! Perhaps a future article will analyze how the team handled pseudo-randomness, a non-zero error rate, and calculating $\alpha$, as I'm sure these problems are just a subtle as the theoretical basis we've discussed here. Regardless, I hope you now feel as though you understand quantum supremacy at a level beyond that of the popular science articles! Despite the simple premise, quantum supremacy requires an incredible range of subjects to completely understand, and the theoretical basis is the first step towards understanding the experiment (and perhaps improving on it in the future). Although the literature is sometimes frustrating to read, I really enjoyed breaking these topics down and completing the derivations, and I hope you enjoyed reading them! And if I've inspired you to start research into quantum computing yourself, I can't wait to see what you come up with! 

\section{Acknowledgments}

A big thanks to Professor Hideo Mabuchi, of the Applied Physics Department at Stanford, for his suggestions and feedback on this paper!

\textbf{Note:} Every effort has been made to ensure the accuracy of the information presented in this article. However, I'm not an expert in quantum information science, so it's entirely possible that one of my explanations was less clear than I expected, or there's an error somewhere. If this is the case, please feel free to contact me at \url{smullane@stanford.edu} so I can update the article accordingly. Thank you! 

\appendix
\appendixpage

\section{PDFs over the continuous random variable $p$}
Here, we'll provide a justification of the following equality:
    \begin{equation}
        \label{eq1}
         \frac{1}{N} \int_{a}^{b} dp' \  \left ( \sum_{p} \delta(p - p') \right ) = \int_{a}^{b} dp \ Ne^{-Np}
    \end{equation}
We'll then use this to motivate another equality that is very useful in ultimately deriving a closed-form expression for the entropy $H(p)$: 
    \begin{equation}
        \label{eq2}
        \frac{1}{N} \int_{0}^{\infty} dp' \ f(p') \sum_{p} \delta(p - p') = \int_{0}^{\infty} dp \ f(p) Ne^{-Np}
    \end{equation}
Let's take a moment to discuss (on a conceptual level) why equation \ref{eq1} is true. The function $Ne^{-Np}$ is a probability density function over the continuous random variable $p$, where $p \in [0, \infty)$. From basic properties of the density function, the integral 
    $$ \int_{a}^{b} dp \ Ne^{-Np} $$
gives the probability that the continuous random variable $p \sim Ne^{-Np}$ is in the range $[a, b]$ for $0 \leq a < b \leq 1$. \textit{An extremely important note: we've restricted our integration range to $[0, 1]$. We'll extend this argument to an arbitrary integration range below!} Equation \ref{eq1}  essentially says that the quantity $(1/N) \cdot \sum_{p} \delta(p - p')$ is an equivalent representation of the probability density function for random variables $p = p(x)$ distributed according to the Porter-Thomas distribution. If this is true, the integral 
    $$ I \equiv \frac{1}{N} \int_{a}^{b} dp' \ \sum_{p} \delta(p - p') $$
must \textit{also} give the probability that the continuous random variable $p$ is in the range $[a, b]$ for $0 \leq a < b \leq 1$. Let's carefully analyze the structure of the integral $I$ to determine if this interpretation is correct. Expanding the sum over all bitstring probabilities $p$ gives a sum of $N$ integrals; the $i$th integral is given by 
    $$ \int_{a}^{b} dp' \ \delta(p_i - p') $$
where $p_i \equiv p(x_i)$ is the probability of obtaining bitstring $x_i$ given some random unitary matrix $U$. If $p_i \in [a, b]$, this integral is equal to 1; otherwise, the integral is 0. Summing all $N$ integrals thus counts the number of bitstrings such that $p$ is contained in the range $[a, b]$. We then divide this count by the total number of bitstrings, which gives the probability that a  selected bitstring will have a probability $p$ in the range $[a, b]$. We already proved that for large $N$, the $p$ are distributed according to the Porter-Thomas distribution, so we can rephrase this as follows: \textit{The integral $I$ gives the probability that the continuous random variable $p \sim Ne^{-Np}$ is contained in the interval $[a, b]$ for $0 \leq a < b \leq 1$.} This is exactly what we wanted to show, up to a point: we've only shown equation \ref{eq1} for integration in the range $[0, 1]$.

\begin{table}
    \centering
    \begin{tabular}{c|c}
        $N$ & $J(N)$ \\
         \hline
        2 & 0.135 \\
        4 & 0.0183 \\
        8 & $3.3 \cdot 10^{-4}$ \\
        16 & $\approx 10^{-7}$ \\ 
        32 & $\approx 10^{-14}$ \\ 
        64 & $\approx 10^{-28}$ \\ 
        128 & $\approx 10^{-56}$ \\ 
        256 & $ \approx 10^{-112}$
    \end{tabular}
    \caption{Value of the integral $J(N)$ for selected values of $N$.}
    \label{pt_ints}
\end{table}

At first glance, it might seem like our equality \textit{can't} hold outside the interval $[0, 1]$. When we consider the integral $I$, the variable $p = p(x)$ is a probability, so all the delta functions $\delta(p_i - p')$ are contained between 0 and 1. So for any $c, d$ such that $1 < c < d$, we must have 
    $$ \frac{1}{N} \int_{c}^{d} dp' \ \sum_{p} \delta(p - p') = 0 $$
If we consider the density function $Ne^{-Np}$, however, we consider $p$ to be a continuous random variable that can be greater than 1. Elementary integration \textit{appears} to show that 
    $$ \int_{c}^{d} dp \ Ne^{-Np} \neq 0 $$
To get around this difficulty, we need to remember that we're working in the limit of \textit{large} $N$. Referencing figure 2, notice that the Porter-Thomas distribution falls to zero very sharply for even a modest $N = 64$. To illustrate just how sharply the distribution goes to zero, let's compute the integral 
    $$ J(N) \equiv \int_{1}^{\infty} dp \ Ne^{-Np} $$
for various values of $N$; the results are given in table \ref{pt_ints}. Obviously, we find that $J(N) \rightarrow 0$ for even the very modest $N \sim 10^2$, so it's reasonable to state that 
    $$ \int_{0}^{1} dp \ Ne^{-Np} \approx \int_{0}^{\infty} dp \ Ne^{-Np} $$
for the values of $N$ that were relevant to the quantum supremacy experiment ($N \approx 10^{15}$). Also note that since $Ne^{-Np} \geq 0$ for all $p$ and $N$, the fact that $J(N) \rightarrow 0$ for large $N$ implies 
    $$ \int_{c}^{d} dp \ Ne^{-Np} \rightarrow 0 \ \text{as} \ N \rightarrow \infty $$
for any $c, d$ such that $1 < c < d$. This allows us to finally state that 
    $$ \frac{1}{N} \int_{a}^{b} dp' \  \left ( \sum_{p} \delta(p - p') \right ) = \int_{a}^{b} dp \ Ne^{-Np} $$
for \textit{any} values of $a$ and $b$ such that $0 \leq a < b$, assuming large $N$. With equation \ref{eq1} in place, we're finally in a position to motivate equation \ref{eq2}. As we noted earlier, equation \ref{eq1} implies that the functions $\alpha(p') \equiv (1/N) \cdot \sum_p \delta (p - p')$ and $\beta(p) \equiv Ne^{-Np}$ are \textit{equivalent representations of the same probability density function}. If $\alpha(p')$ and $\beta(p)$ are equivalent density functions, statistical quantities calculated with these functions should give the same results. In particular, computing the expectation value of a function $f(p)$ over each of these functions should give the same result: 
    $$ \int_{0}^{\infty} dp' \ f(p') \alpha(p') = \int_{0}^{\infty} dp \ f(p) \beta(p) $$
Substituting for $\alpha$ and $\beta$ gives equation \ref{eq2}. Of course, this isn't a proof of equation \ref{eq2}! I instead just wanted to motivate, on a conceptual level, why we should expect equation \ref{eq2} to hold if equation \ref{eq1} is true. It wouldn't make sense to claim that $\alpha$ and $\beta$ are equivalent probability density functions, and then obtain different expectation values from each density function. 


\newpage

\nocite{BLUE}
\bibliographystyle{plain}
\bibliography{ref}

\end{document}